# A framework for systematic analysis of open access journals and its application in software engineering and information systems

Daniel Graziotin, Xiaofeng Wang, Pekka Abrahamsson

Free University of Bozen-Bolzano, Faculty of Computer Science, Piazza Domenicani 3, 39100 Bolzano, Italy
email: daniel.graziotin@unibz.it, phone: +390471016257, fax: +390471016009

## Abstract

This article is a contribution towards an understanding of open access (OA) publishing. It proposes an analysis framework of 18 core attributes, divided into the areas of bibliographic information, activity metrics, economics, accessibility, and predatory issues. The framework has been employed in a systematic analysis of 30 OA journals in software engineering (SE) and information systems (IS), which were selected from among 386 OA journals in Computer Science from the Directory of OA Journals. An analysis was performed on the sample of the journals, to provide an overview of the current situation of OA journals in the fields of SE and IS. The journals were then compared between-group, according to the presence of article processing charges. A within-group analysis was performed on the journals requesting article processing charges from authors, in order to understand what is the value added according to different price ranges. This article offers several contributions. It presents an overview of OA definitions and models. It provides an analysis framework born from the observation of data and the existing literature. It raises the need to study OA in the fields of SE and IS while offering a first analysis. Finally, it provides recommendations to readers of OA journals. This paper highlights several concerns still threatening the adoption of OA publishing in the fields of SE and IS. Among them, it is shown that high article processing charges are not sufficiently justified by the publishers, which often lack transparency and may prevent authors from adopting OA.[1]

---

# Introduction

Transparency and knowledge availability are essential in science. The construction of knowledge is a community-oriented, cooperative and competitive activity (Arunachalam 2008). The scientific method mainly consists of activities related to data. Such activities perform data collection, analysis, publication, critique, and reuse (Molloy 2011). The access to the work of others is necessary in order to evaluate it, to replicate it, and to build upon that knowledge (Crawford & Stucki 1990). Arguably, science is cumulative, no matter if knowledge is added to existing paradigms or if a revolution occurs (Kuhn 1970). Knowledge secures the current scientific beliefs, but also causes paradigm shifts and revolutionizes current beliefs. Scholarly publishing plays an essential role in knowledge sharing because it is essential for the efficiency of research and the dissemination of knowledge (Houghton & Oppenheim 2010).

However, it is known that knowledge is often hidden behind a paywall (Vision 2010) and may not be accessible for many researchers due to the rising subscription costs. As more than 25000 peer-reviewed journals exist and the majority have prices rising (Parks 2002) even faster than inflation (Van Noorden 2013), universities and research centers can only afford a small portion of that total (Harnad et al. 2008). Research opportunities become inevitably lost, and the quality of the research performed is in danger. While the research community is expected to influence the practice by novel research findings related to critical practical problems, the paywall system is known to hinder the knowledge transfer, as companies are reluctant to pay the required subscription costs. As a result, the practice is in danger of basing their decisions solely on the anecdotal material such as blogs and white papers that are freely available on the Internet.

Arguably, a solution for this issue would be to make all research results barrier-free, available on the Internet. Open access is such an initiative (BOAI 2002; Brown et al. 2003; Max Planck Society 2003). Different open access models for publishers exist, namely the golden road, the green road, and the hybrid model. Publishers joining the *golden road* will have all their journal articles freely accessible. Publishers taking the *green road* allow researchers to archive the pre-publication versions of papers on their personal website or in repositories of publications (Harnad et al. 2008). Subscription-based journals may adopt a *hybrid* model between the golden road and the traditional model, where authors pay fees in order to open the electronic versions of articles on an individual basis (e.g., IEEE, 2013). However, these fees are reported to be higher than those required by golden open access publishers.

While the increasing willingness of publishers to allow authors to self-archive the pre-publication versions of articles (Harnad et al. 2008) can elevate the accessibility issue, golden road open access journals would arguably solve it completely. However, despite the potentially noble intention, the golden open access model is also being employed for unethical purposes. There are publishers who exploit the open access scholarly publishing models. These publishers, often called predatory open access publishers, accept a large amount of submitted papers by charging fees, often omitting editorial and publishing services such as peer review (Butler 2012). Authors transfer their copyright to publishers and are subsequently unable to withdraw their papers upon the recognition of fraud (Beall 2012a). While some traditional publishers can be considered predatory as well (Vardi 2012), this study concentrates on open access journals only.

The authors of this article have their research interests in software engineering (SE) and information systems (IS), which are two Computer Science fields and often overlap. Although they are different fields, they have so many commonalities that it is usual for researchers in SE and IS to publish research articles in venues related to both disciplines. It is also common to see program committees and editorial boards of SE and IS research venues composed by people from both fields. In SE and IS literature, the status of open access publishing is rarely researched to the extent of unknown, in spite of some discussions (Boisvert & Davidson 2013). As shown in this study, a limited number of open access journals exist, and the majority of them were born just recently.

In these fields, traditional subscription-based publishers, who employ the hybrid model, advertise open access as an option when publishing an article in their journals. The option of golden road journals is vastly unknown in the fields of SE and IS. Even if a need to explore open access arises, researchers in SE and IS fields are ill-informed when evaluating open access journals as potential venues for their research outcomes. Even worse, they are left to their own devices in choosing which of these journals to target at with their papers.

This paper reports a framework for systematically analyzing open access journals, and its application with 30 open access journals in SE and IS fields, which were selected from among the 386 Computer Science journals of the Directory of Open Access Journals (DOAJ; Bjørnshauge et al. 2013a). The framework extends DOAJ attributes and comprises 18 core attributes, including, among others, the number of journal issues per year, the total number of published articles, the amount of article processing charges and if the fee can be waived, whether a Digital Object Identifier (DOI) is assigned to articles or not, and whether a journal is considered predatory or not. The dataset has been released as Open Data (Graziotin 2013) so that it can be observed, evaluated, employed and extended by other researchers. The analysis was performed under three different viewpoints. First, an analysis of the whole dataset was performed in order to provide an overview of the status of open access journals in the fields. Then, a between-group comparison of journals was performed, according to the presence of article processing charges. Differences between the journals in terms of the framework attributes were highlighted. Lastly, a within-group analysis was performed on the journals requesting publication charges from authors, in order to understand what the added value is according to different price ranges.

This study offers several contributions. First of all, it presents a detailed overview of open access publishing models. Secondly, it raises the need to study open access in the fields of SE and IS while informing the reader of the existence of 30 journals, and it offers a detailed comparison of the journals and it analyzes the issues of open access publishing. Most significantly, it provides an analysis framework born from the observation of data and the literature. Finally, it provides useful recommendations to readers, authors, and publishers of open access articles and journals. This study demonstrates that high publication charges are not sufficiently justified by the publishers, which often lack transparency and may prevent authors from adopting open access.

The rest of the article is organized as follows: in the next section, an overview of the open access definitions and publishing models is provided. Afterwards, an overview is presented on the issues related to finding open access journals and the predatory publishers. The case of open access in the fields of SE and IS is also provided. The *Analysis framework* section describes its areas and how it was constructed. For each of the five framework areas, several attributes are presented and discussed. The *Example application* section describes how the framework was employed to analyze DOAJ listed open access journals in SE and IS. The results are then discussed and recommendations for readers, authors, and publishers are provided. The final section concludes the paper.

## Open access

The term "open access" is associated with the ability to gain access to scientific articles without barriers. Although the term was coined during the last decade, freely accessible scientific journals have been founded since the early 1990s (Björk 2003). Notable examples are New Horizons in Adult Education (1987) and Psycoloquy (1990) (Suber 2009). Since then, 9448 journals were born at the time of this study, according to DOAJ.

Different definitions of open access have been proposed. These definitions are contained in the public statements of Budapest, Bethesda and Berlin. In the Budapest open access Initiative released in 2002, open access has been defined as:

"By 'open access' to this literature, we mean its free availability on the public Internet, permitting any users to read, download, copy, distribute, print, search, or link to the full texts of these articles, crawl them for indexing, pass them as data to software, or use them for any other lawful purpose, without financial, legal, or technical barriers other than those inseparable from gaining access to the internet itself. The only constraint on reproduction and distribution, and the only role for copyright in this domain, should be to give authors control over the integrity of their work and the right to be properly acknowledged and cited" (BOAI 2002).

In the Berlin Declaration on open access to Knowledge in the Sciences and Humanities and the Bethesda Statement on open access Publishing, both released in 2003, a contribution is considered as open access if:

"The author(s) and right holder(s) of such contributions grant(s) to all users a free, irrevocable, worldwide, right of access to, and a license to copy, use, distribute, transmit and display the work publicly and to make and distribute derivative works, […] as well as the right to make small numbers of printed copies for their personal use. A complete version of the work and all supplemental materials, including a copy of the permission as stated above, in an appropriate standard electronic format is deposited (and thus published) in at least one online repository […]" (Brown et al. 2003; Max Planck Society 2003).

The statements of Berlin and Bethesda are almost equal. A difference lies in point 2 of the Bethesda statement: the public, free availability of the stated material is stressed to happen immediately after publication. The similarity in the three definitions is so high that it is common to refer to open access definition as the *BBB[2] definition*, e.g., (Suber 2008).

Recently, the need to distinguish between free access to a research paper and rights to use the resource is suggesting to distinguish between gratis open access and libre open access (Suber 2008). Whereas gratis open access mostly refers to financial barriers in obtaining the scientific papers, libre open access refers to the permissions barriers on how to exploit the already freely accessible resource.

In this paper, we call *open access* journals those whose articles can be found in the peer-reviewed literature and are made freely available on the public Internet without any financial, legal or technical barriers as stated in the BBB definition.

## Publishing models

Three major open access publishing models exist. *Hybrid* open access journals are those where some of the articles are freely accessible without barriers. Upon the payment of a fee by authors, the electronic version of a published paper becomes freely accessible through at least the publisher or the journal portal (e.g., IEEE, 2013).

Publishers joining the *golden road* will have all their journal articles as freely accessible (Harnad et al. 2008). In order to have sustainable publication processes, golden open access publishers usually charge the authors of papers a fee. In this case, the fee is often named *article processing charges*. Other publishers are able to sustain themselves without asking for article processing charges and some authors refer to them as platinum publishers (Beall 2012a). In this paper, we keep the term golden publishers and journals regardless of the presence of article processing charges.

Publishers taking the *green road* allow researchers to self-archive pre-publication versions of papers in their personal website or repositories of publications (Bailey 2008; Harnad et al. 2008). Examples of such repositories are arXiv, figshare, and PeerJ PrePrints. Authors' rights are stated in publishers' copyright transfer forms, but are sometimes difficult to understand. Indeed, it has been estimated that 90% of publishers are green, but only about 10-15% of the papers have been self-archived (Harnad et al. 2008). The SHERPA/RoMEO project is a web system, which lists publishers' copyright agreements and author rights with respect to self-archiving papers (SHERPA 2002).

---

[2] Budapest (February 2002), Bethesda (June 2003), and Berlin (October 2003)

There is a fourth, minor case of open access, where traditional subscription-based journals open their articles after an embargo period. This is often called delayed open access (Suber 2012; Willinsky 2006).

In a simulation of the world switching to open access, it has been estimated that a conversion to the golden road model would bring system savings of about 760 million USD per year in the United Kingdom, whereas a switch to the green road open access model would save about 164 million USD per year (Houghton & Oppenheim 2010). Not only would open access arguably solve the barriers in knowledge acquisition, it would also save significant amounts of funds to be spent in higher education.

## Retrieving open access journals

It is not an easy task for researchers to identify open access journals for their disciplines.

The Open Access Scholarly Publishers Association (OASPA) was founded to guide publishers in setting quality standards for open access (Laakso et al. 2011). The publishers who join the association are required to adhere to a code of conduct (OASPA 2013), which mainly covers ethical issues and transparency. Publishers who fail to adhere to the code of conduct are denied of partnership. Currently, 41 publishers are listed as members of the open access Scholarly Publishers Association. However, the website only lists publishers as members. No journals are listed there, and no disciplines are shown.

The Directory Of open access Journals (DOAJ) (Bjørnshauge et al. 2013a) is a Web-based project aiming to "increase the visibility and ease of use of open access scientific and scholarly journals, thereby promoting their increased usage and impact. The DOAJ aims to be comprehensive and cover all open access scientific and scholarly journals that use a quality control system to guarantee the content." (DOAJ 2013). It allows visitors to browse and search for journals, providing a preliminary overall summary. DOAJ collects golden open access journals of any discipline and presents a preliminary categorization of them. It is the de-facto authority to declare a journal as golden open access and relied on by previous studies, e.g., (Björk & Solomon 2012; Gumpenberger, Ovalle-Perandones, & Gorraiz 2012; Harnad et al. 2008). Currently, a DOAJ entry presents the name of the journal, its ISSN/eISSN, the name of the publisher, the general subject of the journal's scope, the country of provenance of the journal, the language of the publications, the start year of the journal, and whether the journal requires article processing charges or not. However, a DOAJ entry for a journal does not include details such as the amount of article processing charges, the statistics on the published articles, whether the author retains the copyright of the work, and in which databases the journal is indexed. Although useful for the discovery, DOAJ is limited in the evaluation of journals because it only provides limited information about the journals and misses information about the scientific impact of such publication venues (Gumpenberger et al. 2012). However, it is still the best tool available for discovering open access journals.

## Issues and predatory publishers

Unfortunately, open access is not universally seen as the garden of earthly delights in scientific publishing; it has received several critiques, as well. Most notably, the quality of the published articles has worried many researchers (Schroter, Tite, & Smith 2005; Schroter & Tite 2006), and a payment for publication could have a negative impact on the perceived neutrality of peer-review (Worlock 2013). On the other hand, it has been reported that the impact factors of open access journals are about the same level as traditional journals (Pringle 2013) and open access articles are much more accessed than the non-public articles of the same journal (Harnad & Brody 2004).

Some researchers question the sustainability of the open access model (Regazzi 2004; Worlock 2013). An issue threatening the acceptance of golden road journals is the requirement of article processing charges. With average article processing charges between 660.00 USD (Van Noorden 2013) and 900.00 USD (Björk & Solomon 2012), open access reduces the costs of publishing as well as the margin for the publisher (Houghton & Oppenheim 2010; Van Noorden 2013). Article processing charges *per se* are not typically considered something against the open access thinking. Arguably, they are used to cover the maintenance costs, ensuring the longevity of service and the editorial costs, among other expenses. Yet, it is difficult to understand the true costs of scientific publishing, because the subscription prices that libraries pay to publishers are generally hidden by non-disclosure agreements (Van Noorden 2013).

According to Beall (2012b), paid open access may create a conflict of interest, as the more papers are accepted by a journal, the higher is the revenue. Although this holds true for traditional publishers as the revenue comes by selling the access to the papers (Vardi 2012), there are unethical publishers who exploit the open access scholarly publishing models. These publishers attract submission by reporting bogus Impact Factor, or by lying about the number of indexing services and the acceptance rate. These publishers typically accept a large amount of submitted papers by charging high fees, often purposefully omitting the editorial and publishing services such as peer review (Butler 2012). Authors transfer their copyrights to publishers and are subsequently unable to withdraw their papers upon the recognition of fraud (Beall 2012a). Beall (2012b) reports that another damage caused by predatory publishers is that they evoke unethical behavior from scientists, such as plagiarism and not rigorous studies. These publishers notably spam authors, trying to fill editorial positions and looking for research articles (Eysenbach 2008). A documented example is the one reported by Davis (2009). The researcher received several e-mails from the same open access publisher, soliciting submissions of articles to its journals. Davis generated a research article with SCIgen (Stribling, Krohn, & Aguayo 2005), which is a tool by MIT students to generate grammatically correct but nonsense papers in Computer Science. The "nonsense" article was accepted for publication by the publisher after four months, without including any reviewer comments. On the other hand, it must be noted that cases like this happen with traditional publishers, too. For example Mosallahnezhad (2007) got a paper published by the Applied Mathematics and Computation journal. The paper was later withdrawn by the publisher, but a copy has been archived by the SCIgen team and can be reached from SCIgen website together with other examples (Stribling et al. 2005)[3]. Due to the lack of scientific peer-review process, several articles make their way to be used as part of the scientific process of discovery, thus influencing the quality of the subsequent research. In order to improve the transparency of the published articles, some publishers have recently moved to an open pre- and post-peer review model, ranging from the revealing the reviewers' names to the manuscript authors (Smith 1999) to publishing the reviews together with the accepted manuscript – see the journals PeerJ and F1000Research as examples. This process is currently widely unemployed.

As a response to the several issues that he believes are threatening open access publishing, Jeffrey Beall started a collection of "potential, possible or probable predatory scholarly open access publishers" (Jeffrey Beall 2013). The collection, called "Beall's List of Predatory Publishers," currently counts 243 publishers and 126 individual open access journals as predatory. The list and the criteria have been opened to the public and are open to validation, debate, and extension. The employed criteria to determine possible predatory open access publishers are available online (Beall 2013a). They comprise several factors grouped by the categories *Editor and Staff*, *Business Management*, *Integrity*, and *Other*. Examples include: "the journal falsely claims to have an impact factor, or uses some made up measure (e.g. view factor);" "the publisher falsely claims to have its content indexed in legitimate abstracting and indexing services or claims that its content is indexed in resources that are not abstracting and indexing services;" and "evident data exist showing that the editor and/or review board members do not possess academic expertise to reasonably qualify them to be publication gatekeepers in the journal's field". The readers of Beall's list are encouraged to consult the criteria periodically, as it may need time for recently born journals to be assessed. The list has been praised and reported being of high quality and value by Nature (Butler 2012). However, it is important to keep in mind that the list has a *subjective* nature as Jeffrey Beall performs most of the work alone on the list. Therefore, it is expected that the list possesses personal bias. The list has been criticized as well, by both open access advocates and players in the traditional publishing system. For example, it has been reported that several list entries lack a neutral description of what is wrong about a journal (Laika Spoetnik 2011). The employed criteria are difficult to be numerically assessed, yet they are employed in a binary-like fashion; recent open access *megajournals* from respectable traditional publishers may be classified as predatory if a different person evaluates them (Karen 2013). Swoger (2013) argues in *Scientific American Information Culture Blog* that Beall's list should rather be about the quality of the journals than on predatory publishers, and that common sense should be the primary factor when selecting a venue for submitting a study. Vardi (2012), the editor-in-chief of Communications of the ACM, called the list "an informal directory of predatory publishers" and declared that the he resigned from a paid editorial board position of a traditional publisher as soon as he realized that the publishing decisions were made by the publisher rather than the editorial board.  Therefore, predatory publishers can be open access as well as traditional publishers. While the list has generated several arguments for and against its usage, the authors of this study agree with the open access advocate Taylor (2012) that Beall's list is a contribution and *one* solution to predatory publishers and that a better solution has yet to come. However, it is the best option yet[3].

## The case of software engineering and information systems

Software engineering (SE) and information systems (IS) belong to the general discipline of Computer Science. Computer Science is very heterogeneous. In comparison to more established fields, the majority of published papers appear in conference proceedings, and authors have a significant preference for submitting to conferences (Franceschet 2010). Journals still play an inferior role in several Computer Science fields. The Web of Knowledge has very low coverage of journals in these fields (Mattern 2008). It has also been estimated that more than half of Computer Science publications are not ISI-indexed (Wainer, Billa, & Goldenstein 2011). In addition, SE and IS are a small subset of Computer Science research, and open access journals in these fields are even less recognized. As the authors of this article have their research interests in these fields, SE and IS were chosen as the example application of the framework.

Indeed, for SE and IS research, open access is still a new topic (Boisvert & Davidson 2013). It appears that there is opposition to its adoption. The biggest publishers in SE and IS fields are ACM, IEEE, INFORMS, Elsevier, ME Sharpe, Palgrave Macmillan, Springer Science+Business Media, John Wiley and Sons. Seven of them are listed as green road publishers in SHERPA/RoMEO, but each single journal may have different policies. It is unknown how well researchers in the fields are aware of these conditions. These publishers mostly have hybrid open access options, with article processing charges in the order of thousands of dollars. Springer Science+Business

---

[3] The reader should note that this study was conducted months before the publication of Bohannon (2013) "Who's afraid of peer review" in Science, and well before the publication of Beall (2013c) "The Open-Access Movement is Not Really about Open Access". These two articles lent to long, sometimes flaming yet interesting debate between open access advocates, traditional publishers, and librarians. Beall (2013c) caused further disagreement on his list. Both these events are worthy of separate studies. However, this manuscript was already in a soon-to-be-published state when these events happened. Therefore, it does not take into account what these events generated. While none of this changes the results of this study, the authors further acknowledge that Beall's list is a useful contribution but it should be employed with scientific skepticism because of its subjectivity and the recent personal views of its author.

Media is partially embracing the golden road of open access, with the creation of SpringerOpen. By the time this study was conducted, there was only one SE journal with this publisher, but no published articles yet. Researchers in SE and IS need support in order to discover golden open access journals. They have no easy way to obtain this information. Additionally, DOAJ only presents the macro category of Computer Science, which comprised of 386 journals altogether at the time of the study. Sub-categories are currently not available. Researchers in the fields of SE and IS need a better way to compare these open access journals. Several challenges exist to understand the open access phenomenon in SE and IS fields, and there is the need to investigate the open access journals in these fields. A systematic analysis and comparison are the first steps towards making sense of and understanding the compelling attributes of the objects under analysis.

The next section describes a framework for systematically analyzing open access journals, which enables making sense of the jungle of available journals.

## The analysis framework

An objective analysis and comparison of open access journals are a difficult task, like any other comparisons influenced by subjective experiences (Song & Osterweil 1992). There is the need to adopt a systematic methodology and framework to compare the objects under study. The subjective limitations suffered by informal comparison techniques are confined when quasiformal comparisons are employed. Quasiformal comparisons can be approached by different techniques. One technique, which is closer to a traditional scientific method, is to establish a set of critical attributes from various objects and then compare the objects against them (Sol 1983; Song & Osterweil 1992). This is the approach adopted in this study.

The framework has been constructed by the definition of a set of core attributes of the journals under the study. The attributes are divided into five high-level areas, namely, *Bibliographic information*, *Activity metrics*, *Economics*, and *Accessibility*.

The framework attributes have been defined by inspecting the entries of DOAJ, by reviewing and comparing the journals' websites, and by reviewing the existing literature. For each attribute, it is explicated whether it is taken or extended from DOAJ, or it is newly introduced by this study.

## Bibliographic information

This high-level area contains the framework attributes needed for identifying a journal. All these attributes are from DOAJ.

*Name*, *publisher*, *country* – These attributes are needed to identify the journal and to refer to it. In particular, it is necessary to know the publisher of a journal as a way to predict the reliability of the journal itself. Sometimes it is possible to identify a journal as predatory because the publisher is considered predatory. There are independent journals as well. DOAJ includes independent journals but it fills the publisher name anyway, usually with the name of the journal or with the name of the editor-in-chief.

*Electronic ISSN*, *printed ISSN* – the International Standard Serial Number uniquely identifies a journal (Young 1988). When a printed version and an electronic version of a journal are available, two ISSN must be present (ISSN 2013).

Many open access journals are electronic only, as open access was born as part of the digitalization of scholarly articles (Antelman 2004). However, printed journals are better received by the reader and a hybrid model of publication is sometimes preferable (Liu 2006). Although not explicitly distinguished, ISSN and eISSN are part of the DOAJ attributes.

## Activity metrics

This area collects those core attributes, which can be considered as metrics. The attributes have to do with the publications issued by a journal.

*Published Articles, average number of published articles per year, issues per Year* – The reputation and the perceived quality are key factors in authors' decisions to submit to a journal, regardless of its access policy (Schroter & Tite 2006). However, metrics like the Impact Factor and the Immediacy Index are rarely available for young journals. Additionally, in the case of Computer Science research, the ISI Science Citation Index has very low coverage of Computer Science journals (Mattern 2008). On average, 66% of the publications of a computer scientist are not indexed in the Web of Knowledge (Wainer et al. 2011). It has also been shown that ISI Science Citation Index fails to distinguish Computer Science from other areas such as Computational Sciences and Signal Processing (Mattern 2008). This results in severe issues whenever evaluations or rankings in this discipline occur. Similar issues can hold true for other minor disciplines. Lastly, predatory publishers may declare a false Impact Factor. Therefore, traditional metrics are not suitable for a general-purpose analysis framework of open access journals.

Traditional metrics are still a heated debate (e.g. Seglen, 1997), to the point that with open access, a new set of alternative metrics is arising, namely Altmetrics (Priem, Taraborelli, Groth, & Neylon 2010). These metrics aim to measure the impact of the available scientific knowledge through the discussion about the publications happening in social media (Piwowar 2013). Altmetrics are taking place in recently born journals, and it is too soon to evaluate them. They are not available on all open access journals, so they are not employable in this study.

Nevertheless, there is the need to have objective quantitative data for open access journals. It is desirable to help the readers evaluate the establishment and the seriousness of a journal. This can be obtained by observing the number of publications of the journal. Statistics related to the number of publications might indicate if a journal aims to achieve elite status by maintaining a high rejection rate. A low number of publications and a young age might suggest that a journal is still not established. On the other hand, an abnormally high number of publications and a young age may be indicators of predatory open access publishing (Beall 2012a).

## Economics

Open access is not necessarily costless, especially for authors (Houghton & Oppenheim 2010; Suber 2008). This framework area collects those framework attributes necessary to identify the costs to be afforded by authors and their institutions in order to see an accepted manuscript published.

*Article processing charges (USD), avoidable article processing charges* – It is necessary for authors to see whether an accepted manuscript has associated publication charges. As many authors are not ready to face the article processing charges at the moment of publication (Regazzi 2004), the implications for an author's budgets are straightforward. Authors can evaluate the service for the fee they would pay (Van Noorden 2012), given that the average per-article charge for open access publishers is between 660.00 USD (Van Noorden 2013) and 900.00 USD (Björk & Solomon 2012). Some journals have article processing charges that can be waived under certain conditions. For example, there may be grants covering the publication charges for students. DOAJ lists whether a journal has article processing charges or not and whether the fees can be conditionally waived or not.

The employed framework extends DOAJ attributes and quantifies the fees for a research article. It takes into account all the costs that authors will meet in order to have a regular research article published[4]. For example, some journals do not declare publication charges for a paper but ask for "website maintenance fees" instead. This attribute summarizes the total costs to be sustained by authors.

## Accessibility

Accessibility is a core characteristic of open access (Willinsky 2006). It is beneficial to gather data and to describe journals in terms of how much accessibility is given to the published materials.

*Authors keep copyright ownership* – With open access, there is the possibility for authors to retain the copyright of their work. Open access does not imply that authors retain the copyright of their work; therefore, they need to think carefully before transferring all the rights of their work to a publisher (Willinsky 2006). Transferring the copyright to a publisher means that the publisher has the right to prevent the article from becoming open access (Suber 2012, p. 126). It may prevent authors from creating derivative works such as course materials and Web sites (Gutman 2011). To keep the copyright of the work gives the freedom to an author to have it hosted anywhere and to submit a substantially revised version to another venue (where permitted). On the other hand, in the case of plagiarism, the authors have to defend their rights by themselves. Nevertheless, if the publisher owns the copyright of published articles and either takes down an article or ceases its activity, authors have little to do as they do not possess the copyright of their work (see Andreas Holmstrom (2012) and the contained links for examples).

*Provides DOI to articles* – It has been discovered that only 34% of URLs remain operational after a 4-year period (Koehler 2002)[5]. For papers published in electronic journals, it is essential to have persistent and reliable links. A Digital Object Identifier (DOI) is an identifier for online objects of intellectual property and remains the same regardless of the location of the object itself (Langston & Tyler 2004). Many publishers include this feature with their online articles, and it is becoming a necessity with electronic journals (Langston & Tyler 2004).

*Provides a digital preservation mechanism for articles*. Historically, academic libraries received only printed copies of journals. Given the physical existence of the journal copies, the access to knowledge was ensured even after a subscription ended or if a publisher ceased to exist. As academic publishing is migrating to the Web, libraries now have to rent access to digital copies of research articles (Maniatis, Roussopoulos, Giuli, Rosenthal, & Baker 2005).

A difference between information stored on paper and information stored digitally is that the latter "may not be available this time next week" (Reich & Rosenthal 2004, p. 1). This issue is anything but trivial: it is not difficult to find recent reports of disappearing journals and articles, both for traditional subscription-based journals, e.g. (Holmstrom 2012) and for open access journals (Beall 2013b). Thus, it is important for the publishers to ensure a digital preservation mechanism for scientific articles.

*Declared indexing services, verified indexing services* – Online academic citation indexes and databases are a primary source of scientific knowledge as they provide search abilities of manuscripts and the ability to perform citation analysis (Falagas, Pitsouni, Malietzis, & Pappas 2008). The visibility of a manuscript is mostly provided by its presence in key indexes. As previously written, some publishers and journals may inflate the number of indexing services that serve a journal. Therefore, the framework enables the verification of the declared indexing services of a journal.

---

[4] Some journals, for example, have a different price to publish review articles.

[5] As a side note, this manuscript lost two URLs during peer review.

## Predatory issues

The perceived quality of open access has room for improvement (Schroter et al. 2005; Schroter & Tite 2006), and the journal trustworthiness is a key issue to be addressed for open access publishers. The reputation of a journal is the most significant factor for authors to submit (Schroter & Tite 2006); however, open access journals are notably young. They have to be evaluated in terms of perceived trustworthiness. This framework evaluates the journals' trustworthiness in terms of predatory characteristics and general perceived issues.

*Start year* – It is beneficial to know when a journal started its activities in order to understand its longevity. Arguably, it is a potential indicator of the establishment and achievement of the journal and the managing board. The longer an open access journal lasts, the more it reflects the potential of its model to survive (Sotudeh & Horri 2007). If a journal is newly born, it may not be able to self-sustain and establish a network of readers, reviewers and authors.

*Predatory Listed, predatory characteristics and general issues* – As previously reported, there exist unethical publishers and journals, which exploit open access authors only for profit. The employed framework reports a journal as predatory listed if the journal name or the publisher name is listed in Beall (2013a). Additionally, as it may need time for recent journals to be assessed, this framework evaluates each journal using the same criteria (Beall 2013a) and reports the issues.

## Example application: analysis of open access software engineering and information systems journals

As anticipated in the first two sections of this article, there is a huge need to understand open access for the fields of software engineering (SE) and information systems (IS). Given that the authors often submit their manuscripts in both venues, regardless of their official research field, the framework was applied to the sample of all the DOAJ listed journals in SE and IS. This section describes the sampling and analysis methods, and the results of applying the previously described framework.

## Method

To gather data, the authors observed all DOAJ (Bjørnshauge et al. 2013a) journals grouped under the general field of Computer Science. A total of 386 Open-Access journals were retrieved from DOAJ[6] on 2013-05-25. Among the DOAJ entries under the Computer Science category, 30 were retained in the dataset as a result of the following classification method.

Firstly, the resulting dataset was screened for non-relevant entries. The reasons for exclusions from the dataset were: 1) journal off-topic; 2) journal with a wide scope, e.g. including a topic related to SE together with subjects not belonging to the field according to the most recent ACM classification (ACM 2012) such as Bio-Metrics, Nano-Computing; 3) journal with very specialized scope, e.g. only business IS, only for developing countries; 4) journal imposing barriers for submission, e.g. accepts only authors from a certain country; 5) journal website not accessible; 6) journal publishing only proceedings; and 7) journal not supporting the peer review of papers.

---

[6] The query-related URL was http://www.doaj.org/doaj?func=subject&cpId=114&uiLanguage=en

Special attention was given to the declared focus of the journal. In the case that the declared focus was either off-topic or very specialized, the last issue entries of the journal were examined to either confirm or reject the exclusion from the dataset.

The previously reported framework attributes were pre-populated with DOAJ data where possible and subsequently verified; otherwise, the entries were manually gathered by a careful inspection of the journal website and issues. For example, the country of provenance was pre-populated with DOAJ data. The entry was then verified from the journal website first, then from the publisher website if there was one.

Regarding the indexing and abstracting services, each of the reported services were manually examined. If a service was not accessible because of fees, trust was given to the journal. If a service was accessible, the authors tried a lookup of a journal article published at least two years ago. If no results were available, another attempt was made by inputting the journal name. If still no results were available, the reported indexing service was excluded. Other reasons for exclusion were the following: the service did not index any article information (e.g. only the journal name and no metadata on the articles); the service was not an electronic database (e.g. a physical library where no articles can be searched); the service was not an academic abstracting/indexing system at all – e.g. PDF upload services, social bookmarking, and social networks. While it might be tempting to consider excluded services as a result of a lying publisher or journal, it might also be the case that the people behind the publisher or journal did not understand what indexing means.

In order to minimize the validity threats, the data collection procedure was repeated multiple times, with the help of other researchers.

## Results

The results of the systematic analysis are summarized in five tables, each of which is related to a framework area.

Table 1 contains the identification bibliographic aspects of the analyzed journals. The value for the country of origin of the journals is not variegated. 50% of the journals have their publisher's country of origin either in the United States, India, or the United Kingdom. Whereas all the journals possess an electronic ISSN, only 53.3% of the journals possess a "printed" ISSN and are regularly printed on paper.

Table 2 collects data about activity metrics of the journals. On average, the journals are issued 4 times per year (same value for mean and median). Each year, the journals publish a mean of 61.8 articles (standard deviation = 124). The median is 23. Given the standard definition to consider an outlier as any point more than 1.5 times the interquartile range above the third quartile (Navidi 2010), any journal which publishes more than 88.09 articles per year on average can be considered an outlier. 20.0% of the sample falls in this case: IJARCSSE, IJCAIT, IJCISIMA, JC, JS, and JSEA.

Table 3 reports the attributes related to economic issues. 47.6% of the dataset entries do require article processing charges. By taking into account the whole sample, a journal article processing charges is 151.00 USD (standard deviation = 232) on average. The lowest value is 0.00 USD (as it is the median), and the highest value is 800.00 USD.

In Table 4, issues regarding accessibility are provided. 20% of the journals do not require a copyright transfer. Less than half of the journals (46.7%) provide a DOI to the published articles. Regarding digital preservation mechanisms, only 23.3% of the journals declare to employ an archiving system. On average, the journals declare to be indexed by 13.8 services (standard deviation = 17.7) with a median of 7 services per journal. However, the verified services are lower on average: 8.1 for the mean (standard deviation = 8.6) and 5 for the median.

Table 5 collects the framework attributes related to the predatory issues of the journals. Open access journals in the fields of SE and IS are young, as the mean and median for the start year is 2008. Although 13.3% of the journals are in the list of predatory publishers, 76.7% of the dataset entries present issues related to predatory publishing.

It is beneficial to present the dataset with the entries ordered by start year and to cross-examine data in terms of different attributes. A bubble chart is thus drawn which compares the journals in terms of article processing charges, the average number of published articles per year, and whether the journal is in the list of predatory publishers or not. Figure 1 visualizes on the X-axis the timeline between 1993 and 2013 and the amount of article processing charges on the Y-axis. Each circle of the graph has a diameter proportional to the average number of published articles per year; it is black if the journal is in the list of predatory publishers. Figure 1 shows that the number of open access journals in SE and IS grew rapidly after year 2005. It can also be seen that publication charges were not present before 2006 and that journals publishing the higher number of papers per year were born 2006 onwards.

Another viewpoint deals with understanding the journals in terms of the country of origin. Figure 2 shows for each country of provenance the mean of the average number of published articles per year and the average article processing charges. The countries with the highest article processing charges are Egypt, United States, South Korea, and United Kingdom. The countries whose journals have the highest average number of published articles per year are India, United Kingdom, South Korea, and United States. Although inferring conclusions would be misleading, it is still fascinating to note that the journals with the highest article processing charges may be those that publish more articles. Egypt may not show this tendency in Figure 2 because two out of three journals are recent (2009 and 2012).

## Between-group comparison

If the dataset is split into two groups with respect to the presence of article processing charges, the groups are fairly balanced. We call *N-FEE* the group of the journals without article processing charges (53.3% of the dataset) and *FEE* the group of the journals with article processing charges (46.7% of the entries).

Regarding the bibliographic information of the journals, with respect to the country of provenance, the *N-FEE* group does not show particular prevalence, whereas the *FEE* group has the majority of the journals coming from the United States and India (64.3% of the group elements). There is a slight difference between the two groups with respect to the availability of a regular printed version of the journal. 56.2% of the *N-FEE* group has a printed version, whereas 35.7% of the *FEE* group has a printed version.

With respect to activity metrics, there is a difference in the number of issues per year of the two groups. The *N-FEE* group has a median of 2 issues per year, whereas the *FEE* group has a median of 4 issues per year. The two groups show differences in the average number of publications per year. Figure 3 represents the differences between the groups, using boxplots. The *N-FEE* group publishes 18.42 (standard deviation = 13.27, median 13.5) articles per year on average while the *FEE* group publishes 111.64 (standard deviation = 170.43, median = 36) papers on average.

Comparing the groups with respect to economic issues does not make sense, as the *N-FEE* group does not request article-processing charges. However, this aspect will be analyzed in the within-group analysis of the *FEE* group, in the next section.

With respect to accessibility, only 6.2% of the journals in the *N-FEE* group allow authors to keep their copyright; 35.7% of the journals in the *FEE* group, on the other hand, do not require a copyright transfer. 31.2% of the *N-FEE* group provides a DOI to the published articles while 64.3% of the *FEE* group has this feature. Only 6.2% of the *N-FEE* group journals clearly declares a digital preservation mechanism, whereas 42.8% of the *FEE* group declares to provide digital archiving of the manuscripts. There is a difference between the declared indexing and abstracting systems of the two groups. The *N-FEE* group declares an average of 8.81 (standard deviation = 11.73) services, whereas the *FEE* group declares an average of 19.57 (standard deviation = 21.85) services. However, the verified indexed data offers more diversity. For the *N-FEE* group an average of 5.43 (standard deviation = 5.11) services could be verified, whereas for the *FEE* group, an average of 11.07 (standard deviation = 10.85) services could be verified as effective.

With respect to predatory issues, *N-FEE* journals were born earlier than the fee journals, with a median of 2006 for the start year versus a median of 2009 for the *FEE* group. The *N-FEE* group does not have entries listed as predatory in (Beall 2013a) as they do not require article processing charges. However, this analysis shows that about 75% of them present predatory characteristics or general issues related to trustworthiness. On the other hand, 28.6% of the journals in the *FEE* group are considered as predatory publishers in (Beall 2013a) and the analysis of predatory issues reports that 78.6% of them present predatory characteristics or general trustworthiness issues. It should be noted that Beall (and anybody who is collaborating on Beall's list) could be not aware of the existence of some of the journals in this sample. For this reason, Beall's criteria for determining predatory publishers (Beall 2013a) can be employed by the readers, together with common sense and scientific skepticism given their subjectivity.

## Within-group analysis of journals with publication charges
When the journals belonging to the *FEE* group are analyzed further, several observations arise.

The journals from the *FEE* group have article processing charges ranging from 45.00 to 800.00 USD, with an average cost of 321.00 USD (standard deviation = 247.44). It is expected that the publishers will charge the highest amount that the market will allow them. However, it is not clear what justifies this difference in the article processing charges.

IJARCSSE, IJCAIT, GJEIS, and IJSEA belong to the first quartile regarding publication charges. These journals ask for a fee ranging from 45.00 USD to 120.00 USD. GJEIS and IJSEA provide a printed version of the journal. On average, IJARCSSE publishes 648.5 papers per year while IJCAIT publishes 116.0 articles, GJEIS 27.8, and IJSEA 21. All four journals require a copyright transfer from the authors. Only IJSEA provides DOI to the published articles. IJCAIT and IJSEA declare digital preservation mechanisms. Regarding indexing and abstracting, IJCAIT, IJARCSSE, and GJEIS have been verified to be indexed in 2 services while IJICISIMA only in 1. While IJSEA is listed as a predatory publisher in (Beall 2013a), all four journals present predatory factors and general trustworthiness issues.

ISRNSE, ASE, JSEA, and TOSEJ belong to the third quartile of the article processing charges, ranging from 500.00 to 800.00 USD. ASE and JSEA provide a regularly printed version of the journal. On average, JSEA publishes 107.40 articles per year while ISRNSE, ASE, and TOSEJ publish 26.00, 12.75, and 3.00 papers per year respectively. All four journals allow authors to keep the copyright on their published articles. All four journals provide a DOI to the published articles. Only ISRNSE and ASE declare digital preservation mechanisms. Regarding indexing and abstracting, JSEA has been verified to be indexed in 30 services while ASE in 25 services, and TOSEJ, ISRNSE in 5 services. JSEA and TOSEJ are listed as predatory in (Beall 2013a) and present predatory factors and trustworthiness issues from this analysis. On the other hand, ISRNSE and ASE are not considered as predatory and do not present such factors or trustworthiness issues in this analysis.

## Discussion
This section discusses the main results obtained, and aims at offering concrete recommendations for the potential SE and IS authors for open access Journals.

Authors should keep in mind that nearly half (53.3%) of the sampled open access journals are printed on paper regularly. This ratio holds in the *N-FEE* group (the journals without article processing charges), whereas the value falls to 35.7% in the *FEE* group (the journals with article processing charges). While this analysis does not take into account subscription prices for the printed journals, it appears that paying a fee for a published article has nothing to do with the presence of a printed version of a journal.

In this analysis, it has been discovered that open access journals in the sample publish 88.09 articles per year on average (standard deviation = 124). However, 20% of the sampled journals can be considered as outliers as they publish on average more than 1.5 times the interquartile range above the third quartile of such value. While this does not necessarily mean that these journals are either predatory or publish low quality articles, potential authors should keep in mind that 1 in 5 examined journals publish an excessive number of articles and should be carefully evaluated.

At least from this study's sample, there is balance between golden road open access journals requiring article processing charges (53.4%) and those not requiring article processing charges (47.6%). Among those requiring a fee, the requested price is variegated as it ranges from 45.00 USD to 800.00 USD with an average cost of 321.00 USD per article. It can be safely said that for SE and IS, the open access market has a fair choice in terms of price. Although it is expected that the for-profit publishers will charge the highest possible amount of money, it is not clear what justifies this strong variation in the article processing charges. It is not even the case that the journals with the highest article processing charges possess an Impact Factor. Authors should evaluate those journals requiring high article processing charges very carefully.

In terms of accessibility, authors should carefully inspect the journal's website before submission in order to look for copyright transfers. It is not the scope of this article to argue in favor or against copyright transfers to publishers. However, authors have a fair probability to encounter journals that allow them to keep the copyrights on their work. Therefore, attention should be paid to this aspect.

This analysis found that half of the journals (46.7%) provide a DOI to the published articles, but this percentage drops to nearly one third (31.2%) if no article charges are required while it raises to 64.3% if article-processing fees are required. It seems that assigning a DOI to papers raises the publication charges. However, this feature is still provided by *N-FEE* journals and DOI does not have high costs according to CrossRef (2013).

Similar but more alarming is the finding that about three fourths of the journals (76.7%) do not declare a digital preservation mechanism of the papers. That is, it may be that the articles of three fourths of open access journals in SE and IS are in danger of disappearing if the journals lose their content. Only 6.2% of the journals without article processing charges archive the published articles while nearly half of the journals with fees provide such features. This cannot be justified by economic reasons, considering that systems like LOCKSS (Maniatis et al. 2005) are Open Source and even if the required hardware could not be purchased, there are initiatives like CLOCKSS where publishers can archive their content at reasonable prices according to their publishing revenue. For example, for journals with annual publishing revenue up to 250,000 USD, the cost for participating in CLOCKSS is just 200 USD (CLOCKSS 2013). This cost could be sustainable, even for journals without article processing charges, e.g. through donations. All open access publishers should address this issue as soon as possible, as it would raise the trustworthiness of open access publishing.

Authors need to know which agreements have been signed between the journal publisher and the providers of indexing services in order to ensure visibility to accepted papers. As an example, the present authors have gained negative experience in publishing in an open access journal in SE that was deemed to be of sufficient quality but still was not indexed by leading computer science referencing services such as the DBLP (Ley 2002). For a publisher, this means that authors are going to be less likely to submit an article.

It appears that the journals with the lowest article processing charges publish more articles per year on average and provide less indexing services. It is not reassuring, however, that all the sampled journals declare, on the average, 13.8 indexing services while only 8.1 of them could be verified. The delta of the average declared indexes versus the average of the verified indexes is 3.4 for the *N-FEE* group and 8.5 for the *FEE* group. Even though it is not in the scope of this article to assess the quality of the declared indexing and abstracting systems, readers and potential authors are warned that it may be the case that the numbers are inflated by the publishers. Several journals reported a higher and misleading number of indexing and abstracting services. For 26.7% of the journals in this study sample, at least half of the declared services have been removed from the counting because they could not be verified (e.g. the journal articles are not present in the database). Some journals falsely advertise as indexing systems websites such as PDF upload services, social bookmarking websites, social networks, anti-plagiarism services, websites of physical libraries, and websites which only list journal names and some offered services (e.g. SHERPA/RoMEO). Apparently, Google Scholar automatically indexes the articles of all journals. This is not the case with DOAJ and other systems, which need the journal articles to be exposed in certain hierarchical systems and with related metadata. Publishers have at disposal free (Open Source) publishing systems like the Open Journal Systems (Willinsky 2005) and Annotum (Leubsdorf 2012) that provide useful metadata generation to be employed by indexing systems such as DOAJ. Nowadays, missing such features and employing a poorly designed website is obnoxious. When looking for a potential journal, authors are encouraged to observe the declared indexing services and search for the journal entries in them.

Regarding the trustworthiness of the journals, this analysis uncovers new insights. Although Beall's list of predatory journals and publishers (Beall 2013a) contains 13.3% of the sampled journals, this study shows that more than three quarters of the journals (76.7%) present predatory or trustworthiness issues. While this does not necessarily mean that three fourths of the open access journals in SE and IS are predatory, it shows that the majority of the journals cause concerns, which can be perceived through the information provided on their websites. Interestingly, among the four journals with the highest article processing charges, two are not in the predatory list nor do they present issues of trustworthiness, while the other two are included in the predatory list and present several issues discovered by this analysis. Additionally, this analysis shows that almost the same percentage of *N-FEE* and *FEE* journals present predatory issues or trustworthiness issues. For authors, this implies that high article processing charges are not necessarily indicators of predatory publishing, and even more attention must be given while evaluating a journal. As outlined in Figure 1, predatory-considered journals, in black, were all born after 2006. Although four entries is too small a number to infer statistically, we expect more predatory publishers in the coming years. Last but not least, the reader should remember that the authors employed Beall's list and criteria as a tool but remain skeptical because of several validity threats affecting them. The reader is strongly advised to read the last paragraph of the sub-section *Issues and predatory publishers* for more.

This study shows that, in the fields of SE and IS, the open access journals requiring article processing charges do not provide more features than the free-to-submit journals. Nine journals in the *N-FEE* group provide a printed version of the articles. One *N-FEE* journal allows authors to retain copyright. Five *N-FEE* journals provide a DOI to the published articles. Five *N-FEE* journals declare digital preservation mechanisms for the published articles. Seven journals in the *N-FEE* group are indexed in other services besides DOAJ and Google Scholar. It seems that what constitutes a difference between the journals asking for article processing charges with respect to those that do not is the average number of published articles per year and the indexing and abstracting services. A higher number of published articles may suggest that more researchers submit to journals requiring a publication charge. On the other hand, this may also be an indicator of poor editorial and reviewing processes, or even predatory publishing. The journals asking for a fee are indexed and abstracted in a double number of services, with respect to those that do not. Future studies should investigate the quality of these services. Minor indicators of difference are the DOI assignment to articles and the possibility for authors to retain copyrights. Whereas the assignment of a DOI is necessary, it does not justify high publication costs. The possibility for the author to retain the copyright has both advantages and disadvantages.

The added value of publishers requiring article processing charges should be further investigated, but also better explicated by the publishers themselves.

We recommend publishers to describe the publishing process clearly and to justify the publication costs extensively. All the inspected websites fail to justify the costs associated with publication. For example, costs related to the website maintenance, plagiarism checkers, web-hosting fees, and indexing are missing. Given fund availability, there would be no reason to stop a researcher submitting to an open access journal with a high article processing charges if there were a transparent explanation for the demanded money. As many journals without publication charges provide almost the same services to authors, the publishers of journals with publication charges should distinguish themselves in the market by providing notable services. Actually, extremely high article processing charges for a journal may be not justifiable from an author's point of view.

It should also be pointed out that as open access journals are almost entirely Web based, hiring professional web developers and graphic designers would increase the trustworthiness of a journal. Researchers are visitors of the website at first, and then paper authors. If a journal asks money to authors but does not spend money to build a functioning, clear and pleasant website, the authors' perceptions would be negatively affected. The same holds for the published papers: whether pre-formatted word processing templates are given or not, the final published papers should be enjoyable to read.

DOAJ is a valuable tool as a registry of open access journals. However, it was not entirely reliable during the analysis. For example, the journals JAIT, JIC, and JS have the same publisher (Academy Publisher). On DOAJ, the country of provenance for Academy Publisher is reported to be Finland. However, if the publisher's website is carefully observed, it can be seen that the publisher has a regional office in Finland (as well as in China and in North America) and the registered office is in the British Virgin Islands. Fortunately, the errors have always been minor. The authors of this study are pleased to notice that soon after the definition of the analysis framework, DOAJ announced new selection criteria to be employed next year (Bjørnshauge, Brage, Brage, & Jørgensen 2013b), some of which are already contained in this framework.

In addition to carefully reading Beall's list and the criteria to evaluate if a publisher is predatory (Beall 2013a) with a critical eye, we encourage authors to use this systematic analysis together with common sense. For example, one journal - initially analyzed but subsequently excluded from the dataset - was not considered as predatory in Beall's list. However, a careful inspection of the website showed that a misleading impact factor of 1.0 was given without a source; the review process was not explained; the review process length was declared to be of 1-2 weeks with at least 5 reviewers; and the journal published about 100 papers monthly without evidence of dates of submission, revision, and acceptance. This should be enough to consider a journal non-serious. Even more, a quick review of some of the published articles helped to understand that a review process was probably not even performed.

## Conclusions

This article is a contribution towards a better understanding of open access publishing, with a detailed example in the fields of SE and IS. This study empirically demonstrated that high publication charges are not sufficiently justified by the publishers in these fields, which often lack transparency and may prevent authors from adopting open access. It showed that there are no features provided by journals with article processing charges which are not offered by those not requiring charges from authors. The article warned authors to investigate which agreements have been signed by the journal publisher in order to ensure visibility to accepted papers. It also raised important concerns, e.g., the articles of three fourths of open access journals in SE and IS may be in danger of disappearing if the journals lose their content. Last but not least, this study showed that open access journals and publishers in the fields of SE and IS have a significant margin of improvement regarding the perceived trustworthiness. It was shown that more than three fourths of the journals present predatory or general trustworthiness issues according to Beall's criteria. While Beall's list and criteria possess several validity threats, the biggest one being the subjectivity of the list, it is the only tool available still. The authors and readers of open access journals can employ it with a critical, skeptical, and scientific eye. Common sense should be the best tool for evaluating open access and subscription-based journals.

The implications of this study are useful for authors, especially regarding predatory publishers. However, publishers are those who should benefit more from this study. Open access publishers in the fields of SE and IS should aim for higher transparency and improve the quality of the journals websites and related services.

It is not within the scope of this systematic analysis to declare a winner among open access journals, neither is it within its scope to bring shame to specific journals. However, this study made easy for the reader to use the analysis, the comparisons, the reported criteria, and the recommendations in order to identify unethical journals and publishers.

Embracing open access in SE and IS is currently a challenge for researchers, as it is difficult to understand the phenomenon let alone to evaluate it. Preconception towards open access can be pulled down with advocacy but that might not be sufficient for SE and IS research. Open access in these fields is threatened by several issues which can be easily solved by the publishers. This paper aims to highlight such concerns with the hope that in the near future, open access can be studied as a real alternative to traditional publishing systems for the SE and IS fields.

## Acknowledgments

The authors are thankful to Elena Borgogno for her valuable help during the study and when writing this article. The authors would like to thank Christian Gumpenberger for the insightful comments he offered to improve the manuscript. Lastly, the authors are thankful to two anonymous reviewers for the several suggestions that significantly improved the article.

# Tables

**Table 1 Bibliographic information of the analysis framework.**

| Journal abbreviation | Name | Publisher | Country | Electronic ISSN | Printed ISSN |
|---|---|---|---|---|---|
| AJIS | Australasian Journal of Information Systems | Australasian Association for Information Systems and Australian Computer Society | Australia | 1449-8618 | No |
| ASE | Advances In Software Engineering | Hindawi Publishing Corporation | Egypt | 1687-8655 | 1687-8663 |
| CSIS | Computer Science and Information Systems | Comsis Consortium | Serbia | 1820-0214 | 1820-0214 |
| eISEJ | e-Informatica Software Engineering Journal | Wroclaw University of Technology | Poland | 1897-7979 | 2084-4840 |
| EJISE | Electronic Journal of Information Systems Evaluation | Academic Conferences Limited | United Kingdom | 1566-6379 | No |
| GJEIS | Global Journal of Enterprise Information Systems | Academic Management Society | India | 0975-1432 | 0975-153X |
| IJARCSSE | International Journal of Advanced Research in Computer Science and Software Engineering | S.S. Mishra | India | 2277-128X | No |
| IJCAIT | International Journal of Computer Applications & Information Technology | Mahadev Educational Society | India | 2278-7720 | No |
| IJCISIMA | International Journal of Computer Information Systems and Industrial Management Applications | Machine Intelligence Research Labs (MIR Labs) | United States | 2150-7988 | No |

| IJIEEB | International Journal of Information Engineering and Electronic Business | MECS Publisher | Hong Kong | 2074-9023 | 2074-9031 |
|---|---|---|---|---|---|
| IJISTE | International Journal of Information Systems and Telecommunication Engineering | HyperSciences Publisher | Tunisia | 1737-9245 | 1737-9237 |
| IJSE | International Journal of Software Engineering | Software Engineering Competence Center (SECC) | Egypt | 2090-1801 | 1687-6954 |
| IJSEA | International Journal of Software Engineering & Applications (IJSEA) | Academy & Industry Research Collaboration Center (AIRCC) | India | 0975-9018 | 0976-2221 |
| IJSEIA | International Journal of Software Engineering and Its Applications | SERSC | South Korea | 1738-9984 | No |
| ISRNSE | ISRN Software Engineering | Hindawi Publishing Corporation | Egypt | 2090-7680 | No |
| JAIT | Journal of Advances in Information Technology | Academy Publisher | United Kingdom | 1798-2340 | 1798-2340 |
| JC | Journal of Computers | Academy Publisher | United Kingdom | 1796-203X | 1796-203X |
| JIA | Journal of Information Architecture | Research & Education Group in IA (REG-iA) | Denmark | 1903-7260 | No |
| JICT | Journal of Information and Communication Technology | UUM Press | Malaysia | 1675-414X | 2180-3862 |
| JIITO | Journal of Information, Information Technology, and Organizations | Informing Science Institute | United States | 1557-1319 | 1557-1327 |
| JIOS | Journal of Information and Organizational Sciences | University of Zagreb | Croatia | 1846-9418 | 1846-3312 |
| JRPIT | Journal of Research and Practice in Information Technology | Australian Computer Society | Australia | 1443-458X | No |
| JS | Journal of Software | Academy Publisher | United Kingdom | 1796-217X | No |
| JSEA | Journal of Software Engineering and Applications | Scientific Research Publishing | United States | 1945-3124 | 1945-3116 |
| JSI | Journal of Systems Integration | Czech Society of Systems Integration | Czech Republic | 1804-2724 | No |
| SEAIJ | Software Engineering: An International Journal | Delhi Technological University | India | 2249-9342 | No |
| TIJSCSE | The International Journal of Soft Computing and Software Engineering | Advance Academic Publisher | United States | 2251-7545 | No |
| TISJMIS | The International Scientific Journal of Management Information Systems | University of Novi Sad | Serbia | 1452-774X | No |
| TOISJ | The Open Information Systems Journal | Bentham open | United States | 1874-1339 | No |
| TOSEJ | The Open Software Engineering Journal | Bentham open | United States | 1874-107X | No |

**Table 2** Activity metrics of the analysis framework.

| Journal abbreviation | Published articles | Average number of published articles per year | Issues per year |
|---|---|---|---|
| AJIS | 474 | 47,40 | 1 |
| ASE | 51 | 12,75 | 1 |
| CSIS | 337 | 37,44 | 4 |
| eISEJ | 54 | 9,00 | 1 |
| EJISE | 136 | 8,50 | 2 |
| GJEIS | 111 | 27,75 | 4 |
| IJARCSSE | 1297 | 648,50 | 12 |
| IJCAIT | 116 | 116,00 | 6 |
| IJCISIMA | 391 | 97,75 | 1 |
| IJIEEB | 94 | 23,50 | 4 |
| IJISTE | 11 | 3,67 | 6 |
| IJSE | 66 | 13,20 | 2 |
| IJSEA | 63 | 21,00 | 4 |
| IJSEIA | 252 | 42,00 | 4 |
| ISRNSE | 26 | 26,00 | 1 |
| JAIT | 111 | 37,00 | 4 |
| JC | 1658 | 236,86 | 6 |
| JIA | 20 | 5,00 | 2 |
| JICT | 82 | 7,45 | 1 |
| JIITO | 46 | 6,57 | 1 |
| JIOS | 155 | 22,14 | 2 |
| JRPIT | 220 | 22,00 | 4 |
| JS | 1320 | 188,57 | 12 |
| JSEA | 537 | 107,40 | 12 |
| JSI | 83 | 27,67 | 4 |
| SEAIJ | 21 | 10,50 | 2 |
| TIJSCSE | 60 | 30,00 | 12 |

| TISJMIS | 96 | 13,71 | 4 |
|---------|-----|-------|---|
| TOISJ | 15 | 2,50 | 1 |
| TOSEJ | 18 | 3,00 | 1 |

**Table 3 Economics of the analysis framework.**

| Journal abbreviation | Article processing charges USD | Avoidable article processing charges |
|---|---:|---:|
| AJIS | 0 | Yes |
| ASE | 600 | No |
| CSIS | 0 | Yes |
| eISEJ | 0 | Yes |
| EJISE | 0 | Yes |
| GJEIS | 100 | Yes |
| IJARCSSE | 45 | No |
| IJCAIT | 50 | No |
| IJCISIMA | 130 | Yes |
| IJIEEB | 0 | Yes |
| IJISTE | 0 | Yes |
| IJSE | 0 | Yes |
| IJSEA | 120 | No |
| IJSEIA | 330 | No |
| ISRNSE | 500 | No |
| JAIT | 0 | Yes |
| JC | 360 | No |
| JIA | 0 | Yes |
| JICT | 0 | Yes |
| JIITO | 0 | Yes |
| JIOS | 0 | Yes |
| JRPIT | 0 | Yes |
| JS | 360 | No |
| JSEA | 700 | No |
| JSI | 0 | Yes |
| SEAIJ | 0 | Yes |
| TIJSCSE | 190 | Yes |

| TISJMIS | 0 | Yes |
|---------|------|-----|
| TOISJ | 250 | No |
| TOSEJ | 800 | No |

**Table 4 Accessibility of the analysis framework.**

| Journal abbreviation | Authors keep copyright | Provides DOI to articles | Digital preservation mechanisms | Declared index | Verified index |
|---|---|---|---|---|---|
| AJIS | No | Yes | No | 0 | 1 |
| ASE | Yes | Yes | Yes | 25 | 25 |
| CSIS | No | Yes | No | 13 | 13 |
| eISEJ | No | Yes | No | 6 | 5 |
| EJISE | No | No | No | 2 | 2 |
| GJEIS | No | No | No | 0 | 2 |
| IJARCSSE | No | No | No | 22 | 2 |
| IJCAIT | No | No | Yes | 2 | 2 |
| IJCISIMA | No | No | No | 1 | 2 |
| IJIEEB | No | Yes | No | 16 | 12 |
| IJISTE | No | No | No | 8 | 3 |
| IJSE | No | No | No | 2 | 3 |
| IJSEA | No | Yes | Yes | 16 | 8 |
| IJSEIA | No | No | No | 7 | 7 |
| ISRNSE | Yes | Yes | Yes | 5 | 5 |
| JAIT | No | Yes | Yes | 44 | 19 |
| JC | No | Yes | Yes | 54 | 27 |
| JIA | No | No | No | 0 | 1 |
| JICT | No | No | No | 2 | 2 |
| JIITO | No | No | No | 8 | 6 |
| JIOS | No | No | No | 9 | 7 |
| JRPIT | No | No | No | 26 | 6 |
| JS | No | Yes | Yes | 54 | 27 |
| JSEA | Yes | Yes | No | 64 | 30 |
| JSI | Yes | No | No | 0 | 2 |
| SEAIJ | No | No | No | 3 | 2 |
| TIJSCSE | No | Yes | No | 10 | 7 |
| TISJMIS | No | No | No | 2 | 3 |
| TOISJ | Yes | Yes | No | 7 | 6 |
| TOSEJ | Yes | Yes | No | 7 | 5 |

**Table 5 Predatory factors and general issues of the analysis framework.**

| Journal abbreviation | Start year | Predatory listed | Predatory factors and issues |
|---|---|---|---|
| AJIS | 2003 | No | No |
| ASE | 2009 | No | No |
| CSIS | 2004 | No | No |
| eISEJ | 2007 | No | No |
| EJISE | 1997 | No | Editor contact is difficult to be found; review process lacks the details |
| GJEIS | 2009 | No | The Editorial Board comes from the same country; publishing process completely missing; offers fast review of 2 weeks for 50.00 USD; review process is unknown |
| IJARCSSE | 2011 | No | The Editorial Board comes from the same country; contact e-mail is a general e-mail domain; 5 days of peer review declared; review process is not described; declares an IF of 2.080, no evidence; |
| IJCAIT | 2012 | No | The Editorial Board comes from the same country; Editor-in-chief cannot be identified; peer review process not clear; contact e-mail is general e-mail domain; |
| IJCISIMA | 2009 | No | Peer review process omits details; scope not clearly stated; |
| IJIEEB | 2009 | No | Editorial Board does not list a single contact information; could not find information on editor-in-chief |
| IJISTE | 2010 | No | Editorial Board does not list a single contact information; instruction for authors are common for all the publisher's journals; not all declared indexed services are indexing services of articles; whole journal website consists in 5 web pages |
| IJSE | 2008 | No | Editorial Board does not list a single contact information; could not find information on editor-in-chief; instructions for authors are two bullet points; only given contact info is general "executive secretary"; whole journal website consists in 5 web pages |
| IJSEA | 2010 | Yes | Editorial Board does not list a single contact information; could not find information on editor-in-chief; revisions due only 3 days after notification; advertises PDF upload systems as indexing services; only given contact info is a general e-mail |
| IJSEIA | 2007 | No | No |
| ISRNSE | 2012 | No | No |
| JAIT | 2010 | No | Editorial Board only lists a general e-mail as contact info; no fees guaranteed for 2013; Only half of the declared indexing systems could be verified; Some declared indexes are not indexing services or only index the journal name |
| JC | 2006 | No | Only half of the declared indexing systems could be verified; Some declared indexes are not indexing services or only index the journal name |
| JIA | 2009 | No | Editorial Board does not list a single contact information; Last issue was Spring 2011; Contact page not working; missing author guidelines; |
| JICT | 2002 | No | The Editorial Board comes from the same country and universities; review process not described |
| JIITO | 2006 | No | The Editorial Board boards currently has only 5 people |
| JIOS | 2006 | No | The Editorial Board comes from the same country and university (however, the university is also the publisher); readers need to register to read papers; some declared indexing services do not exist anymore |
| JRPIT | 2003 | No | Many declared indexes are in reality part of a single one; Review process is stated but difficult to be reached |
| JS | 2006 | No | Editorial Board only lists a general e-mail as contact info; only half of the declared indexing systems could be verified; some declared indexes are not indexing services or only index the journal name |
| JSEA | 2008 | Yes | Review process not clear, declares notification within 4 weeks; Additional 50 USD cost for each page above 10; Many indexing systems are part of a single service; Not all indexing systems index abstracts / articles; Not all indexing systems are indexing systems |
| JSI | 2010 | No | The Editorial Board almost comes from the same country; review process lacks clarity |
| SEAIJ | 2011 | No | No |
| TIJSCSE | 2011 | No | Charges 250 USD for a "fast track" mode, no other information provided; Some claimed indexing services are not paper indexing services |
| TISJMIS | 2006 | No | Asks for 60 USD for a 6 week standard review (not clear if extra or not); 150 USD for a 3 weeks "fast track" mode; |

| | | | Publication fees are 450.00 USD for 6-10 pages length; 650 USD for 15 pages limit; 850 USD for 20 pages limit; Review process stated but vague; |
|---|---|---|---|
| TOISJ | 2007 | Yes | Editorial Board entries are incomplete, in the form <first name initial><last name>(<country>) and lack contact information; aims and Scope is brief (1-line sentence); peer review process lacks details; "endorsements" (testimonials) page looks fake and the entries never mention the publisher's journals; |
| TOSEJ | 2007 | Yes | Editorial Board entries are incomplete, in the form <first name initial><last name>(<country>) and lack contact information; aims and Scope is brief (1-line sentence); peer review process lacks details; "endorsements" (testimonials) page looks fake and the entries never mention the publisher's journals; |

## Figures

**Figure 1 Start year vs. article processing charges vs. average number of published articles per year of open access journals.**

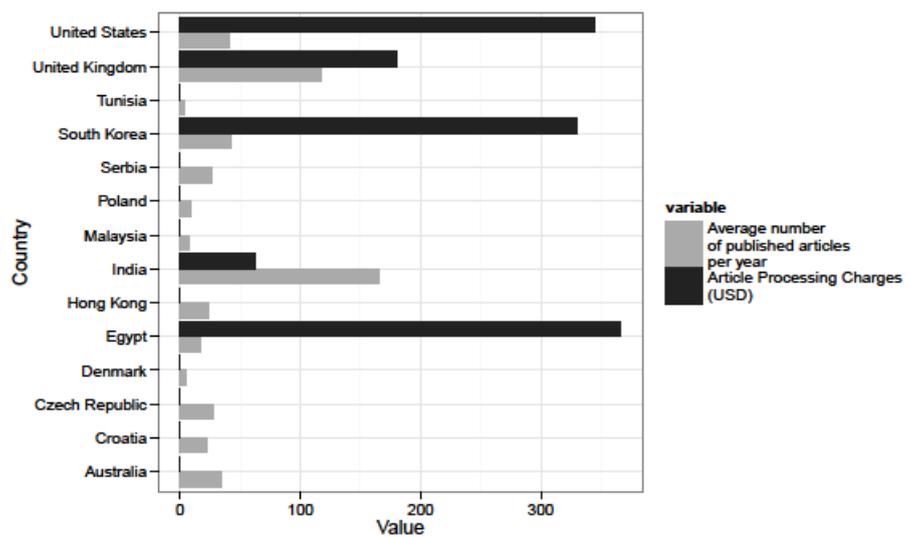

**Figure 2** Average article processing charges and mean of the average published articles per year of open access journals, by country of provenance.

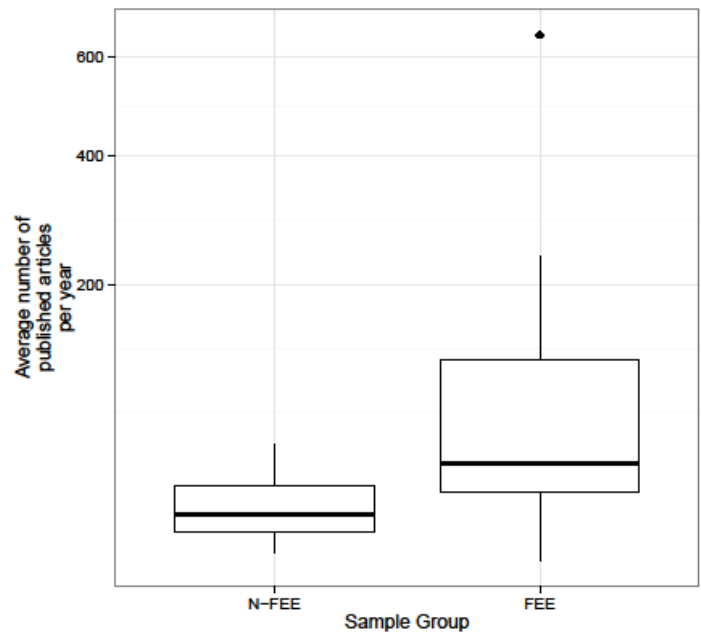

**Figure 3 Average number of published articles per year for the N-FEE and the FEE groups of open access journals.**